# Influence of visual cues on head and eye movements during listening tasks in multi-talker audiovisual environments with animated characters

*Maartje M. E. Hendrikse[1], Gerard Llorach[1,2], Giso Grimm[1] and Volker Hohmann[1,2]*

[1]Medizinische Physik and Cluster of Excellence 'Hearing4all', Universität Oldenburg, Germany.
[2]Hörzentrum Oldenburg GmbH, Oldenburg, Germany

**Abstract**

Recent studies of hearing aid benefits indicate that head movement behavior influences performance. To systematically assess these effects, movement behavior must be measured in realistic communication conditions. For this, the use of virtual audiovisual environments with animated characters as visual stimuli has been proposed. It is unclear, however, how these animations influence the head- and eye-movement behavior of subjects. Here, two listening tasks were carried out with a group of 14 young normal hearing subjects to investigate the influence of visual cues on head- and eye-movement behavior; on combined localization and speech intelligibility task performance; as well as on perceived speech intelligibility, perceived listening effort and the general impression of the audiovisual environments. Animated characters with different lip-syncing and gaze patterns were compared to an audio-only condition and to a video of real persons. Results show that movement behavior, task performance, and perception were all influenced by visual cues. The movement behavior of young normal hearing listeners in animation conditions with lip-syncing was similar to that in the video condition. These results in young normal hearing listeners are a first step towards using the animated characters to assess the influence of head movement behavior on hearing aid performance.

## 1. Introduction

Established procedures for hearing-aid evaluation in the laboratory use stationary acoustic-only conditions with fixed source and receiver positions (e.g., Luts et al. 2010). These procedures do not reflect natural dynamic communication conditions, which might explain why directional microphones perform well in the stationary laboratory conditions, but not in real life (Bentler 2005; Cord et al. 2004). Furthermore, recent achievements in hearing-aid development result in algorithms that interact with or rely on head- and eye-movement behavior (e.g., Tessendorf et al. 2011; Abdipour et al. 2015). To properly evaluate these novel algorithms, realistic head- and eye movement is required, and it is assumed that this can only be achieved by including visual cues in the test environment. The long-term goal of our research is therefore to create a laboratory-based, audio-visual test environment in which ecologically valid movement behavior and ecologically valid task performance can be measured under comfortable conditions for both normal-hearing listeners (to provide reference data) and hearing-aid users. The ecological validity of a measure indicates that it is meaningful under real life conditions. As a first step towards this goal, this study measured the influence of visual cues on head- and eye-movement behavior and looked at the influence on task performance and perception. Young, normal-hearing listeners were included in this study to develop and evaluate the methods, and to provide reference movement data for later studies involving older subjects and subjects with hearing aids.

One approach to investigate movement behavior experimentally is to perform recordings in the field (Tessendorf et al. 2012). However, in the field it is difficult to achieve reproducible test conditions and to systematically assess the factors underlying movement behavior. An alternative approach is to use virtual environments that offer high-quality, plausible sound-field simulations in the laboratory (e.g., Grimm, Kollmeier, et al. 2016; Cubick & Dau 2016). Using virtual environments in the laboratory makes experiments more realistic and closer to field tests, but sufficiently controlled to be reproducible. However, when compared to clinical tests, the higher complexity and variability of the virtual environments make certain experiments less sensitive to small effects on, e.g., speech-reception thresholds. Whereas the sound reproduction in laboratory virtual environments has been evaluated extensively, also for hearing aids (e.g., Grimm, Kollmeier, et al. 2016; Cubick & Dau 2016), it is not yet clear to what extent movement behavior in such environments is ecologically valid.

Presenting visual stimuli makes the test environment more realistic. It may also influence the listener's head- and eye-movement behavior. Furthermore, it is well known that seeing the speaker[1] can improve speech intelligibility (Sumby & Pollack 1954). Speech intelligibility improves because of lip reading (e.g., Erber, 1969; Macleod & Summerfield, 1987) and seeing the movements of the speaker's mouth and face, which focuses listeners' attention on specific spectro-temporal locations in the speech waveform, enhancing sensitivity to the acoustic information. This results in an additional increase in speech intelligibility that is different from lip reading (Grant 2001; Schwartz et al. 2004). Gestures made by the speaker also help to improve speech intelligibility (Munhall et al. 2004; Drijvers & Özyürek 2017). Visual information is also important for social aspects, such as gaze direction for conversational turn-taking (Vertegaal et al. 2001), and for using facial expressions to judge emotions (Ekman & Oster 1979).

The use of animations as visual stimuli permits more freedom when creating different real-life scenarios and avoids making time-consuming video recordings (involving related privacy issues). Animations are finding their way into audiological research (Meister et al. 2016), and a number of aspects need to be considered. It is known that both the appearance and behavior of the animated characters are important for the perceived realism and that those two need to be in balance, as higher realism of the appearance may lead to higher expectations of the behavioral realism (Slater & Steed 2002). Furthermore, it is important to stay away from the so called "Uncanny Valley" (Mori 1970), where a too human-like appearance can cause feelings of unease when there is also an abnormal feature (Seyama & Nagayama 2007). As mentioned above, behavioral features such as mouth movements, gestures, gaze direction and facial expression can increase speech intelligibility for human speakers and should therefore also be included in animated characters to achieve speech perception similar to that for real speakers. Studies of these behavioral features in virtual characters usually focus on their effect on speech intelligibility (e.g., Fagel & Clemens 2004; Meister et al. 2016), or on the perceived realism by the user (e.g., Lee et al. 2002; Le et al. 2012). However, it remains unclear how these behavioral features affect the user's movement.

---

[1] In this paper the term 'speaker' is used to refer to persons or animated characters in the tasks, otherwise the term 'loudspeaker' is used.

This study focuses on the effect of different visual cues from animated characters on head- and eye movement, on combined speech intelligibility and localization task performance, and on perceived speech intelligibility, perceived listening effort and the general impression of the audiovisual environments. It seeks a level of realism of the animated characters that results in ecologically valid movement behavior. In the laboratory, using video recordings of real people, either 2D recordings, stereoscopic recordings or 3D recordings with light-field cameras (Akeley 2012), is the most realistic condition that can reproducibly be achieved. In this study, 2D video recordings of real people projected onto a cylindrical screen were used to compare to the animations. Animation conditions that induce movement behavior similar to the video condition were considered more ecologically valid. Furthermore, a condition in which the speakers cannot be seen (audio-only) was included as an additional reference. Two important features of the animated characters, lip-syncing and gaze direction, were analyzed in detail to determine which information is necessary to measure ecologically valid movement behavior and the task performance of the listener. Not only were concordant lip-syncing (speech-driven) and gaze direction (towards active speaker) tested, but also discordant lip-syncing ('fish mouth') and gaze direction (randomized), to determine how these features influence movement behavior, task performance, and perception.

To formulate a hypothesis on the effect of visual cues on movement behavior in challenging listening environments, information is needed about strategies that could drive movement behavior. In multi-talker conversations, most of the time (62%) people look at the individual they listen to (Vertegaal et al. 2001). This is probably due to social factors and because of the information that can be obtained by looking at the speaker's mouth movements, gestures, gaze direction and facial expression. Thus looking at the active speaker could be one strategy to increase speech intelligibility. Another strategy would be to move the head to maximize the signal-to-noise ratio (SNR). Studies have shown that although hearing-impaired listeners successfully increase SNR or speech level by head movements, young normal-hearing listeners have difficulty spontaneously finding a beneficial head orientation, and half of them do not move at all (Brimijoin et al. 2012; Grange & Culling 2016). The results of the aforementioned study were obtained without visual information, and it is not clear which effect visual information has on orienting behavior. A third possible strategy could be one driven by localization. It is known that target localization is important for understanding the target speech (Yost et al. 1996) and that head movements are important for localization (Wallach 1939). Kim et al. (2013) studied how people move during a localization task and found that when localizing, people turn their head towards the sources. Of course this strategy is mainly important if the active speaker cannot be seen, otherwise localization is trivial. The first two strategies are often conflicting, because the optimal head direction can be too far away from the target direction to see the target speaker's face. Both strategies usually require some movement, which is important for localization. Based on these movement strategies, it is expected that people will look at the active speaker if the speaker is visible. However, if there is discordant or no lip-syncing, this strategy will provide no speech intelligibility benefit and people are more likely to follow the SNR-optimization strategy. If the speakers are not visible, people might turn their head toward the active speaker when following the localization strategy, or turn their head to optimize SNR.

Therefore, the first hypothesis of the current study is that the animation conditions with speech-driven lip-syncing induce movement behavior similar to that in the video condition. As a second hypothesis, the task performance is expected to increase when adding lip-syncing and gaze direction towards the active speaker. Perception of the different visual conditions will be influenced by how realistic the visual conditions are. Therefore, as a final hypothesis, the most realistic animation condition, with speech-driven lip-syncing and gaze direction towards the active speaker, is expected to receive the best subjective ratings of the animation conditions, and it is expected that subjects feel comfortable in this condition. Note that, since there is no comparison to the video condition for the speech intelligibility and localization task (see section 2), the ecological validity of the task performance will not be assessed. It can however be determined if, as stated in the second hypothesis, the task performance in the animation conditions increases compared to the task performance in the audio-only condition.

## 2. Method

Movement behavior, task performance and perception were tested and evaluated using three tasks. First, a listening task was carried out (Task 1). The aim of this task was to measure natural movement behavior. Subjects were asked to listen to a conversation between four persons in cafeteria background noise and subsequently answer multiple-choice questions about the content. Second, a German adaptation of the coordinate response measure (CRM) task (Moore 1981; Bolia et al. 2000) (Task 2) was used to measure combined speech intelligibility and localization performance, as well as movement behavior. Finally, to measure how subjects perceived the different visual conditions, they were asked to compare and rate the visual conditions from the first listening task using a questionnaire (Task 3). The conditions used in the tests included an audio-only condition in which the speakers could not be seen, a video condition with recordings of real people, and conditions using animated characters. The animated characters could have different lip-syncing: concordant lip-syncing (speech-driven), discordant lip-syncing ('fish mouth'), or no lip-syncing. In addition, there were different gaze patterns: concordant (towards the active/target speaker), discordant (randomized), or static (always towards the subject). Note that the same set of audio files was used for all conditions. For Task 1, the audio-only, video, and animated conditions with different lip-syncing and gaze to the subject were used. The effect of gaze direction was also studied by adding the condition of gaze to the target and speech-driven lip-syncing. For Task 2, no video material was available, so this task was only tested with the 'audio-only' and animated conditions. In terms of measurement time and concentration, Task 2 was more demanding on the subjects for one condition, reducing the number of conditions that could be measured for this task. The focus was placed on gaze direction of the animated characters, because analyzing the contribution of the speech-driven lip-syncing to speech intelligibility was not the aim of the current study and the study design would not have been suitable for an in-depth analysis. For Task 3, the same visual conditions were used as for Task 1. See Table 1 for an overview.

Table 1: Overview of visual conditions tested for the listening to conversations task (1), CRM task (2) and rating task (3)

| Task 1: Listening to conversations<br>Task 3: Subjective rating | | Task 2: Speech intelligibility and localization |
|---|---|---|
| Audio-only | | Audio-only |
| Video | | |
| Animated | No lip-sync, gaze to subject | |
| Animated | Lip-sync, gaze to subject | Lip-sync, gaze to subject |
| Animated | Lip-sync, gaze to target | Lip-sync, gaze to target |
| Animated | | Lip-sync, gaze random |
| Animated | Fish mouth, gaze to subject | |

## 2.1. Participants

Tests were carried out with 14 self-reported normal-hearing participants (seven male, seven female) aged between 19 and 35 years, most of them students of Oldenburg University. Subjects had normal or corrected-to-normal vision. The subjects were shown a video that introduced them to the different visual conditions and that explained that head- and eye movements were measured. They were asked to behave as they would in everyday life. The subjects gave written consent. All experiments were approved by the ethics committee of Oldenburg University.

## 2.2. Visual stimuli

The video conversations were recorded in the same physical location in which the experiments were carried out, a circular space with uniform background, using two cameras (Canon EOS 700D) and one microphone (Neumann KM 184) for each speaker. The video recordings were split into four different videos, one for each speaker, so they could be rendered as planes in the 3D environment at the desired positions. The virtual characters were created with MakeHuman (MakeHuman_Team 2016) using the recorded persons as a reference (Fig. 1), and imported into the 3D game engine Blender (version 2.78a, Roosendaal 1995).

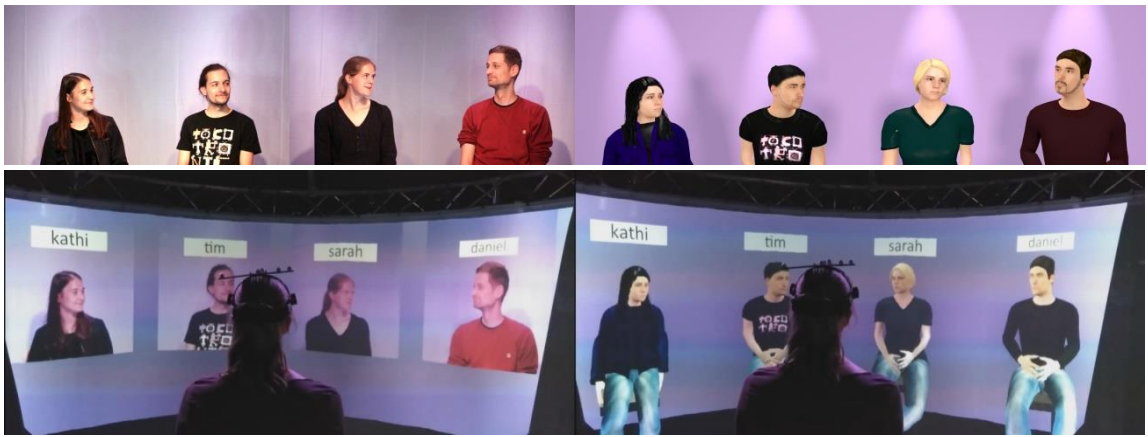

**Figure 1:** Picture of the actors (top left) and the animated characters (top right) and of the video of the actors (bottom left) and animated characters (bottom right) being presented to a subject.

The animated characters could move their head and eyes through inverse kinematics, following a 3D point in space. Inverse kinematics calculates the joint angles necessary to reach or point to a 3D location in the scene. In our case, it was implemented using tools provided by Blender ('Track To' Bone Constraint), in which the angles of the head and eyes were automatically calculated. When the eyes pointed at a 3D point in the scene, the total movement was a combination of eye- and head rotation, as shown in Figure 2. Head-eye angle relationships were made to look natural and are in correspondence with values found in the literature. A study by Uemura et al. (1980) reports that for gaze directions of 50° and 10°, the percentage of the angle obtained by head movement is about 62% and 93%, respectively. When interpolating, this would result in head rotations of 13.3° and 29.6° for gaze directions of 15° and 45°, respectively. The virtual characters blinked every 0.1-4 seconds with blink durations of 0.15-0.5 seconds. The rules for the blink timings were taken from Ruhland et al. (2014; 2015).

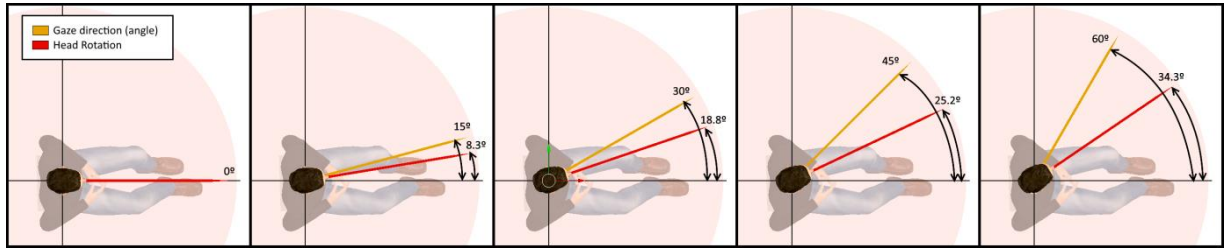

**Figure 2:** The head rotations (red line) of an animated character for a number of gaze directions (orange line). This illustrates that the total movement is a combination of eye and head rotation, for example in the second panel from the left the head rotation is 8.3° and the eye rotation is 6.7°, resulting in a gaze direction of 15°.

The speech-driven lip-syncing shaped the mouths of the animated characters as a weighted average between three blend shapes, where the weights were based on the estimated energy of the formants (Llorach et al. 2016). The 'fish mouth' lip-syncing was a simple sinusoidal open-and-close movement of the mouths with frequencies ranging from 0.5 to 2 Hz when the animated characters were talking (the frequency of the sinusoidal movement was kept constant during speech but was different across speech utterances). It therefore provided no information other than which speaker was talking. No lip-syncing means that the mouths of the animated characters were always closed. The gaze direction 'gaze to subject' means that the animated characters were looking at the subject the whole time. The exact meaning of 'gaze to target' depends on the task. For Task 1, the animated characters that were not speaking were looking at the active speaker. The active speaker was looking in the same direction as before he started speaking, thus either at the subject or at another animated character. For Task 2, all animated characters were speaking simultaneously and they were looking at the target speaker, who was looking at the subject. In the 'gaze random' condition, each animated character was looking in a different direction; they either looked at one of the other animated characters or at some point next to the subject. The size of the videos and virtual characters was adjusted to the size of real humans at the distance of the projection screen from the subject, which was 1.76m.

## 2.3. Setup

In the laboratory, a 16-loudspeaker (Genelec 8020B) horizontal ring array at ear level was used to present the audio (loudspeaker n positioned at $11.25° + (n-1) \times 22.5°$ azimuth from frontal direction) (Fig. 3). The loudspeakers were controlled by TASCAR (versions 0.128-0.142, Grimm et al. 2015), which used horizontal 7th Order Ambisonics panning with max-rE decoding (Daniel et al. 1998). To present the visual stimuli, a projector (NEC U321H) was used, which projected the images from a Blender scene (version 2.78a, Roosendaal 1995) onto a cylindrical, acoustically transparent, screen of 3.52m diameter, achieving a 120° field of view (Fig. 3). Before the image was sent to the projector, the warping necessary for projecting onto a cylindrical screen was enabled by the rendering pipeline of the game engine Blender. The equipment was attached to a cloth-covered metal frame that reduced environmental sounds, light and room reflections (Fig. 3). Head movement (rotation in the horizontal plane) of the subjects was measured with an infrared camera (TrackIR 5 by Naturalpoint), which tracked six reflective markers on a custom-made cap worn by the subjects, using a sample rate of 120 Hz (Fig. 3). The head-movement sensor had an angle-dependent error due to the slightly tilted camera axis; the RMS value of this error was 1.3°. The noise of the head-movement sensor was below 0.1°. Eye movement (angle relative to the head in the horizontal plane) was measured with a custom-made wireless electrooculogram (EOG) amplifier and a sample rate of 50 Hz. The EOG amplifier consisted of a high-impedance operational amplifier, a 10bit analog-digital converter, and a Bluetooth transmitter. A built-in, first-order high-pass filter compensated for the electrode voltage drift. The angle relative to the head in the horizontal plane was measured by the EOG sensor with an accuracy of roughly ±10°. This is accurate enough to determine which of the speakers the subject was looking at. The head tracker and EOG sensors were calibrated by displaying on the cylindrical screen a cross at the currently measured head direction while the subject was seated in front of it. Subjects were then asked to adjust the cap until they felt the cross matched the direction they faced. For the EOG sensor calibration, the cross was then moved to the left/right of the currently measured head direction by a known number of degrees and the subjects were asked to follow the cross with their eyes. A camera pointed at the subjects was used to record their faces during the listening task (Fig. 3). TASCAR and LabStreamingLayer (Medine 2016) were used for time synchronization and data logging.

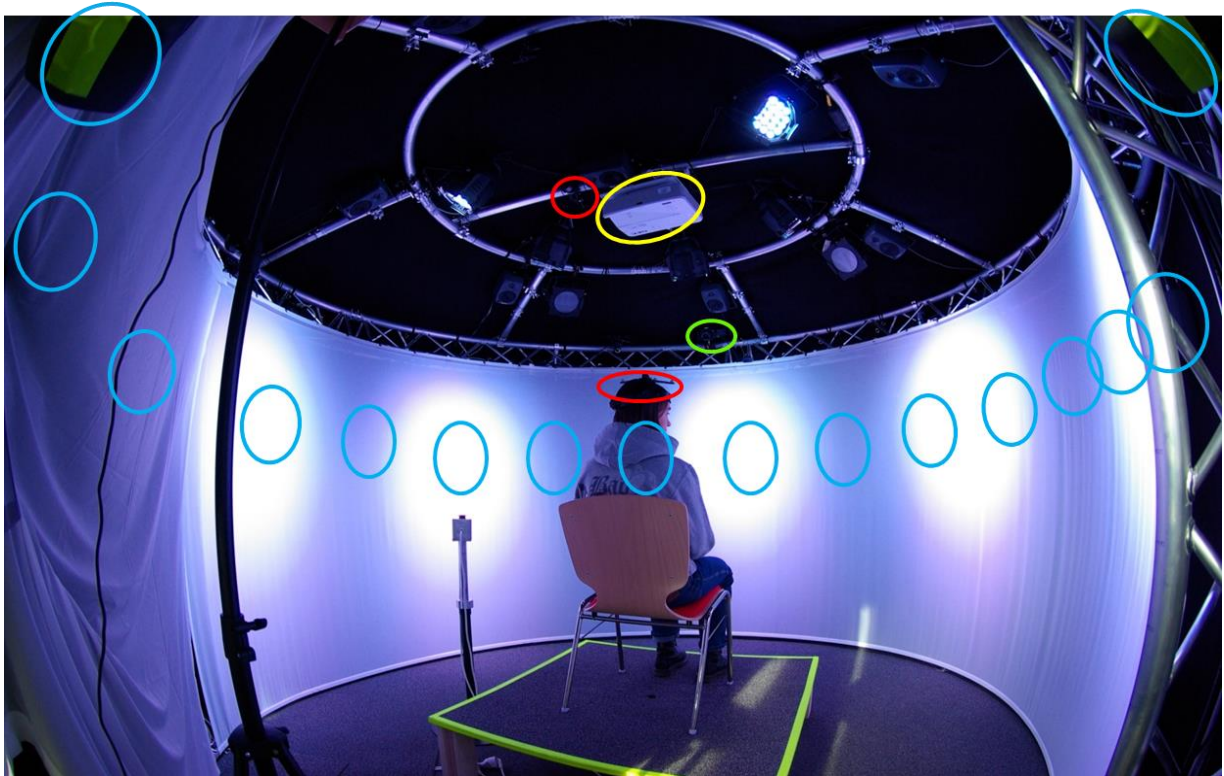

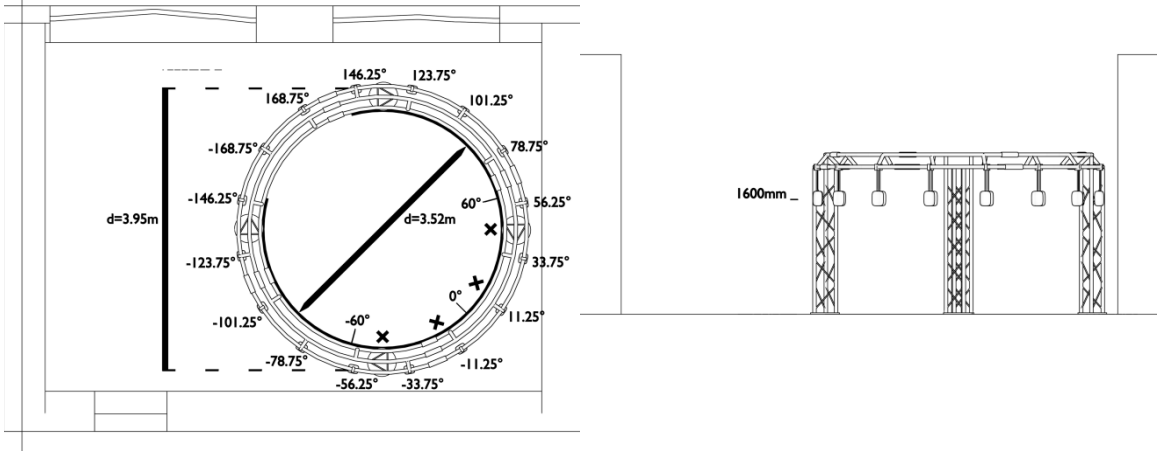

**Figure 3:** Setup. Top: photo (fish-eye optic) of the setup showing the loudspeakers (blue ellipses), projector (yellow ellipse), TrackIR sensor and marker cap (red ellipses) and camera (green ellipse). Bottom left: schematic top view of the setup and the room, with the dimensions and loudspeaker positions (crosses) indicated. Bottom right: schematic side view of the setup and the room.

## 2.4. Task 1: Listening to conversations

Seven different conversations between 4 actors were recorded and played to the subjects. The script of the conversations was based on common conversation topics (food, holidays/traveling, weather, work, plans, movies, anecdotes), as listed in the FOLK database (Deppermann & Hartung 2011). Each conversation took approximately 1.5 minutes (1min 24s to 1min 39s). The actors performed the conversation freely, to ensure natural behavior of the actors in the recordings. The videos of the conversations with both real actors and animated characters were published online (DOI: 10.5281/zenodo.1203198). The conversations were played in diffuse cafeteria background noise (first-order Ambisonics recording from a real cafeteria, spatially upsampled to $7^{th}$ order using a method similar to that of Zotter et al. (2014)). The level of the cafeteria noise was on average 58.4

dB(A). The average levels for the individual speakers in the conversations were: 56.2±1.0 dB(A) and 54.8±1.4 dB(A) for the female speakers; 57.2±1.5 dB(A) and 57.5±1.7 dB(A) for the male speakers. Natural speech was used and therefore small differences in speech level could not be avoided. The sound levels resulted in an average SNR of -2.0 dB(A) (-2.8 dB(A) to -1.3 dB(A)) or a more meaningful average segmental SNR after Quackenbush et al. (1988) of about -6.5 dB (-6.9 dB to -5.7 dB). The conversations could be played as video, as audio-only or as animation. The actors in the video and the animated characters were positioned at -45°, -15°, +15° and +45° and did not change position. The task for the subjects was to listen to the conversation and then to answer multiple-choice questions about the content. The multiple-choice questions were declared before the experiment to make sure that the subjects were paying attention. The answers were, however, not analyzed. From the seven conversations, one was always used to familiarize and train the subjects with the task played in the video condition. The other conversations were played in one of the conditions mentioned in Table 1. Which conversation was played in which condition, as well as the order of conditions, was randomized. To check how accurately the subjects could follow the conversational turn-taking, a guided movement behavior condition was added, in which the subjects were asked to closely follow the active speaker with their eyes, but they were not prompted about their head-movement behavior. This was carried out by 11 of the subjects in the 'audio-only', video, and 'lip-sync, gaze to target' visual conditions, with the same conversations and positions of the speakers as used in the free-movement conditions. The guided-movement behavior condition was measured after the second task, to make sure it did not influence the movement behavior of the subjects.

## 2.5. Task 2: Speech intelligibility and localization

A German adaptation of the Coordinate Response Measure (CRM) task (Moore 1981; Bolia et al. 2000) was used to simultaneously measure localization and speech intelligibility. In this task, 4 speakers positioned at -45°, -15°, +15° and +45° simultaneously voiced a sentence of the form: "Los geht's *(Goethe / Lessing / Heine / Schiller)* entscheide dich für *(Blau / Rot / Grün / Weiss)* und *(Zwei / Drei / Vier / Fünf)*." In English: "Let's go *name* choose *color* and *number.*" The subjects had to attend to the speaker saying "Goethe" and respond with the position of this speaker as well as the color and number said by that speaker. Audio recordings of one male and one female speaker were available (Schepers et al. 2017). One additional female and one additional male speaker were created by changing the formant frequencies, pitch, pitch range and duration of the audio recordings with PRAAT (Boersma & Weenink 2017). Changing the duration made sure all speakers had a slightly different speech rate and word-onset timing. The original male voice was changed to an artificial female voice by multiplying the formants by a factor of 1.1, the pitch by 1.8, the pitch range by 1.1 and the duration by 0.95. The original female voice was changed to an artificial male voice by multiplying the formants by a factor of 1/1.1, the pitch by 1/1.8, the pitch range by 0.9 and the duration by 0.95. When doing these transformations, the voice rhythm and naturalness of the original voices were maintained, although a native speaker could recognize that the voices were artificial. During the task, these differences could no longer be noticed, as all speakers were talking simultaneously. The subjects showed a similar task performance for the artificial and original voices (see Section 3.2.), which suggests that there was no influence of the artificial voices. Each voice was matched with one animated character and this combination remained the same. The presentation levels for the individual speakers were 61.1±1.1 dB(A), resulting in an overall sound level of 67.5 dB(A). Each speaker acted twice as target speaker at each location, which resulted in 32 trials for each visual condition. The order and the key words were randomized, with the constraint that the key words voiced by the target speaker were always unique. Animated characters with different patterns of gaze directions were compared for this task (Table 1). The different visual conditions were explained to the subjects, and it was mentioned that in one condition, the animated characters were looking at the target speaker, but not which condition this was. The order of the visual conditions was randomized among the subjects.

## 2.6. Task 3: Subjective rating of the visual conditions

The subjects were asked to compare and rate the different visual conditions from Task 1 (Table 1). On a touchscreen, they could switch between the different visual conditions while listening to and looking at the conversation from the training, as previously displayed on the cylindrical screen. On each change of condition, audio and video were faded out within 0.5 seconds, and faded in to the new condition within 0.5 seconds, to avoid an effect of animation transitions on the rating. Each condition could only be rated after being selected and displayed for 8s. Subjects were asked to rate the conditions by general impression, perceived speech intelligibility, perceived listening effort and naturalness of the audio. To rate the general impression, the question "How do you like/enjoy the virtual experience?" was asked. To rate the perceived speech intelligibility, the question "How much of the conversation did you understand?" was asked. To rate the perceived listening effort, the question "How much effort does it take to follow the conversation?" was asked. For the question about listening effort, the same German version of the labeled categories was used as used by Krueger et al. (2017a; 2017b). To check whether the visual condition affected the perceived naturalness of the audio, the question "How natural does the acoustic environment sound?" was asked. The subjects could answer on a continuous scale with five or seven labeled categories (see Fig. 8), and entered their answers using a slider on a touchscreen.

## 2.7. Data analysis and measures

**EOG data processing.** The built-in high-pass filter of the EOG device could not adequately compensate for a drift in the data when a subject was looking at the same angle for a longer period of time. Therefore, a different filter was applied. The cutoff frequency of the built-in high-pass filter was estimated by applying a first-order Butterworth high-pass filter to the location of the active speaker over time for a particular conversation and comparing this to the measured gaze direction for the subjects in the guided movement behavior condition for that same conversation. The cutoff frequency estimated by this method was 0.025 Hz. The filter was inverted by swapping the transfer function coefficients and this inverted filter was applied to the data to obtain the original data with drift. After compensating for the built-in high-pass filter of the EOG device, the EOG data were extrapolated with a linear fit using linear regression and filtered with a moving average filter with a time constant of 27s to remove the drift. The angle presented in the calibration versus the measured potential was fitted using linear regression to obtain the slope and offset needed to convert the EOG data from millivolt to degrees. For some subjects one of the EOG electrodes had become loose during the measurement, causing instances of electrode saturation when the electrode lost contact. Data points measured during such saturation were removed, as well as data points up to 5 s after saturation, because the drift compensation took some time to adjust. For the EOG data, on average 2.2±5.3% of the data for Task 1 and 3.3±6.1% of the data for Task 2 was removed. In the worst case (one subject in one condition) 28.1% of the EOG data was removed for Task 1 and 31.4% for Task 2, with a maximum time gap between two consecutive data points of 17.3 s for Task 1 and 23.1 s for Task 2.

**Head-tracker data processing.** Missing or incorrect data points in the head movement data could occur when one or more of the reflective markers on the cap were occluded or out of range of the sensor. An offline processing algorithm was used that could salvage some of these data points by fitting the marker configuration of the cap to the measured reflections using the known distances between the markers and recalculating the rotation in azimuth. Some errors could still occur if the fitting was erroneous, causing sudden angle jumps in the head-movement data. Data points after such angle jumps up to the next angle jump were removed if the angle jump was larger than 6°. For smaller angle jumps, it was not possible to reliably determine which data points were correct, so no data points were removed in such cases. After processing, on average 0.5±2.3% of the head movement data was missing for Task 1 and 1.4±3.0% for Task 2. In the worst case (one subject in one condition) 20.3% of the head movement data was missing for Task 1 and 12.9% for Task 2, with a maximum time gap between two consecutive data points of 7.0 s for Task 1 and 18.7 s for Task 2.

**Handling missing data.** In Task 1, the conversations had durations of about 1.5 minutes, so even with 28.1% of the EOG data missing there were enough data points remaining to calculate the measures stated below. Likewise, in Task 2, 64 trials were measured, so even with 31.4% of the EOG data missing there were enough data points left to make the calculations. Missing data points were replaced by not-a-number, so that they did not affect the calculation of the mean and standard deviation.

**Measures Task 1.** The head movement data was resampled with the EOG sampling rate and added to the EOG data to obtain the gaze direction of the subjects. For Task 1, the root-mean-squared value (RMS) of the difference between the gaze direction and the active speaker position was calculated as a measure of how accurately subjects looked at the active speaker ('**gaze-direction error**'). The time ranges that were within 0.5s after a speaker change were excluded for this measure. Furthermore, the **number of gaze jumps** (difference of more than 7° between adjacent data points, corresponding to an angular velocity of more than 350°/s) and the '**gaze delay**', the delay between a speaker change and the next gaze jump in the correct direction (with a maximum of 1.2s), were calculated. The number of gaze jumps was included as a measure of how often subjects shifted their center of gaze (to a different speaker). These gaze jumps are therefore large saccades with an angular velocity of more than 350°/s, corresponding to the value of around 500°/s for saccades of 30° (Ruhland et al. 2014). Finally, the **SNR perceived by each subject** for their individual head movement was approximated with a simulation. Using TASCAR, the in-ear signals of a virtual listener making the head movements of each subject were recorded using the HRTF database of a Brüel & Kjaer head and torso simulator (Thiemann et al. 2015). The long-term average segmental SNR of these recordings was calculated, and the value for the better ear selected in each segment using the method described in Quackenbush et al. (1988).

**Measures Task 2.** For Task 2, the **head- and eye-movement trajectories** during each trial were calculated relative to the position of the target speaker. Trials where location, color, and number were correct were separated from those where one or more answers were incorrect. The average was calculated for each visual condition, and for correct and incorrect trials separately. Also, for the correctly-answered trials, the **head direction relative to the center during the number key word** (at the end of the trial) was calculated and compared to the head direction with maximal SNR. This was done separately for the center and edge speakers. The sign for the data of the negative target speaker locations was flipped and the data were pooled. The head direction with maximal SNR was predicted using TASCAR and the HRTF database. Furthermore, the **average number of gaze jumps** **during each trial** and the **percentage of trials in which location, color, and number were correct** were calculated.

## 3. Results

### 3.1. Task 1: Listening to conversations

The movement behavior of the subjects during this task was analyzed by calculating the gaze-direction error, the number of gaze jumps, the gaze delay and the SNR for the individual head orientations, as described in section 2.6. First, the different visual conditions were compared for the free movement behavior. The gaze-direction error, number of gaze jumps, gaze delay and SNR for the individual head orientations for the free movement behavior are plotted in Figure 4.

**Statistics for effect of visual condition on free movement behavior.** A repeated-measures ANOVA of the results for the free movement behavior (values in Table 2) shows that there is a significant effect of the visual condition on the gaze-direction error, on the number of gaze jumps, on the gaze delay, and on the SNR for the individual head orientations. Contrasts between the video condition and the other conditions revealed that in the 'audio-only' condition and 'no lip-sync, gaze to subject' conditions, the subjects had a significantly larger gaze-direction error than in the video condition (Fig. 4A). In the 'audio-only' condition, subjects also made significantly fewer gaze jumps (Fig. 4B), had a significantly larger gaze delay (Fig. 4C), and the individual head movements of the subjects resulted in a significantly higher SNR (Fig. 4D) compared to the video condition.

**Explanation for effect of visual condition on free-movement behavior.** In the 'audio-only' condition, most subjects did not move at all and always pointed their head and eyes towards the center (0°). This behavior explains the significantly larger gaze-direction error, the significantly smaller number of gaze jumps, the significantly larger gaze delay and also the significantly higher SNR compared to the video condition. Always looking at the center would also result in a higher SNR, because one of the ears is pointed more towards the target source than it would be when looking at the target. For all other animation conditions except for 'no lip-sync, gaze to subject', there were no significant differences in movement behavior compared to the video condition. In the 'no lip-sync, gaze to subject' condition, the gaze-direction error was also significantly higher than in the other animation and video conditions, but not as high as in the 'audio-only' condition. Furthermore, the delay of gaze jumps after a speaker change was similar to that in the other animation and video conditions. This indicates that subjects were looking at the active speaker, but for a smaller percentage of the time.

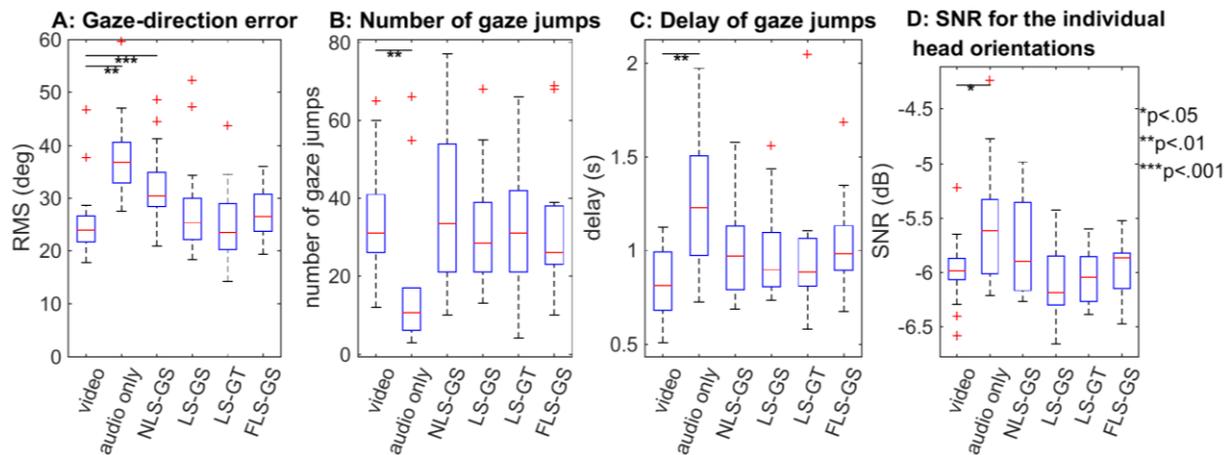

**Figure 4** Comparison of free-movement behavior measured in the different visual conditions for the listening to conversations task (N=14). The visual conditions are indicated in the figure as: video, audio-only, NLS-GS (no lip-sync, gaze to subject), LS-GS (lip-sync, gaze to subject), LG-GT (lip-sync, gaze to target), FLS-GS (fish mouth, gaze to subject). Boxplots (range, interquartile and median, outliers are plotted as red crosses) are of the following movement behavior measures: A. Gaze-direction error (RMS error of the gaze direction relative to the active speaker); B. Number of gaze jumps; C. Delay of gaze jump after a speaker change; D. SNR for the individual head orientations.

Moreover, a comparison was made between the guided-movement behavior and free-movement behavior conditions. The gaze-direction error, gaze delay and SNR for the individual head orientations for both guided- and free-movement behavior are plotted in Figure 5 for all subjects that performed the guided-movement task (N=11).

The number of gaze jumps made by the subjects in the guided-movement behavior condition should be equal to the number of speaker changes. The average number of gaze jumps the subjects made in the guided- movement condition was about 19, whereas the number of speaker changes in the conversations was 11 to 19. The correspondence between the average number of gaze jumps and the number of speaker changes is an indication that the subjects were following the instructions correctly.

**Effect of visual condition on guided-movement behavior.** A factorial repeated-measures ANOVA (values in Table 2) of the results in Figure 5 shows that the visual condition had a significant effect on the gaze-direction error and on the gaze delay, but not on the SNR for the individual head orientations. These significant effects are for the combined data of the guided- and free-movement behavior conditions. When looking at differences between visual conditions for the guided-movement condition only, simple effects reveal no significant differences between visual conditions for the gaze-direction error, but there were significant differences between the visual conditions for the gaze delay. In the 'audio-only' condition, subjects had a larger gaze delay. Apparently the visual information in the video and 'lip-sync, gaze to target' conditions helped to more rapidly identify the new active speaker after a speaker change.

**Comparison of free- and guided-movement behavior.** Factorial repeated-measures ANOVA (values in Table 2) also showed that the test condition (free- or guided-movement behavior) had a significant effect on the gaze-direction error and on the SNR for the individual head orientations, but not on the gaze delay. Analysis of simple effects helps to identify those visual conditions for which there was a difference between the free- and guided-movement behavior. It revealed a significantly larger gaze-direction error in the free-movement behavior condition than in the guided-movement condition for all visual conditions ('audio-only', video and 'lip-sync, gaze to target'; Fig. 5A). The SNR for the individual head orientations was significantly higher in the free-movement behavior condition than in the guided-movement condition for the audio-only and video conditions (Fig. 5C). Clearly, always looking at the active speaker did not result in the best SNR.

In short, the guided movement behavior showed that without visual information ('audio-only' condition) it took longer to identify the new active speaker after a speaker change, but this did not affect the overall accuracy with which subjects looked at the active speaker. It also showed that in the free-movement behavior condition, subjects did not look as accurately at the active speaker as when specifically instructed to do so (larger gaze-direction error) and that always looking at the active speaker did not result in the best SNR.

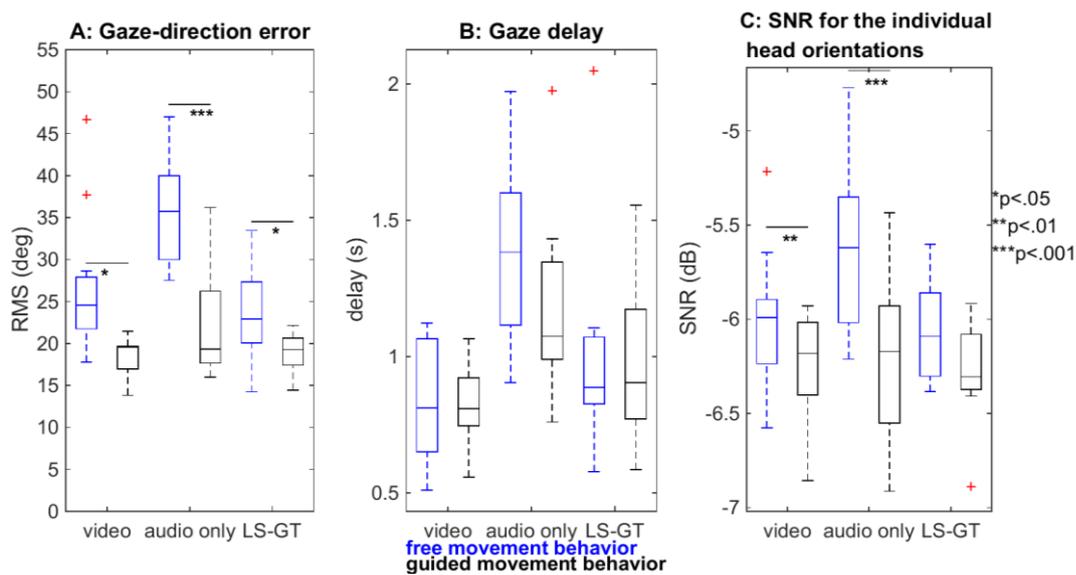

**Figure 5** Comparison of free-movement behavior (blue, subset of data in Fig. 5) and guided-movement behavior (black) for the listening to conversations task (N=11). Boxplot (range, interquartile and median, outliers are plotted as red crosses) of movement behavior in the 'audio-only', video and 'lip-sync, gaze to target' (LS-GT) visual conditions for: A. Gaze-direction error (RMS error of the gaze direction relative to the active speaker); B. Gaze delay (delay of gaze jumps after a speaker change); C. estimated SNR for the individual head orientations.

**Table 2 Statistics results Task 1.**

| Measure | Movement behavior | Comparison | F | p | Effect size |
|---|---|---|---|---|---|
| gaze-direction error | free-movement behavior | main effect of visual condition | F(5,65) = 11.09 | < .001 | |
| | | 'audio-only' vs. video | F(1, 13) = 22.82 | < .001 | partial $\eta^2$ = .64 |
| | | 'no lip-sync, gaze to subject' vs. video | F(1, 13) = 18.00 | < .01 | partial $\eta^2$ = .58 |
| | guided- and free-movement behavior combined | main effect of visual condition | F(2, 20) = 15.41 | < .001 | |
| | | main effect of test condition | F(1, 10) = 46.07 | < .001 | |
| | | guided vs. free for 'audio-only' | F(1, 10) = 75.18 | < .001 | r = .94 |
| | | guided vs. free for | F(1, 10) = 8.14 | < .05 | r = .67 |

|  |  | video |  |  |  |
|  |  | guided vs. free for 'lip-sync, gaze to target' | F(1, 10) = 10.14 | < .05 | r = .71 |
| number of gaze jumps | free-movement behavior | main effect of visual condition | F(3.12, 40.52) = 6.46* | < .001 |  |
|  |  | *Mauchly's test was significant: $\chi^2$ = 31.23, p < .01, correction with Greenhouse-Geisser estimate of sphericity $\varepsilon$ = .64* |  |  |  |
|  |  | 'audio-only' vs. video | F(1, 13) = 10.71 | < .01 | partial $\eta^2$ = .45 |
| gaze delay | free-movement behavior | main effect of visual condition | F(5, 65) = 4.58 | < .01 |  |
|  |  | 'audio-only' vs. video | F(1, 13) = 15.69 | < .01 | partial $\eta^2$ = .55 |
|  | guided- and free-movement behavior combined | main effect of visual condition | F(2, 20) = 17.57 | < .001 |  |
|  |  | guided-movement behavior: video vs. 'audio-only' vs. 'lip-sync, gaze to target' | F(2, 20) = 3.85 | < .05 | r = .53 |
| SNR for the individual head orientations | free-movement behavior | main effect of visual condition | F(3.06, 39.72) = 5.24* | < .05 |  |
|  |  | *Mauchly's test was significant: $\chi^2$ = 29.53, p < .05, correction with Greenhouse-Geisser estimate of sphericity $\varepsilon$ = .61* |  |  |  |
|  |  | 'audio-only' vs. video | F(1, 13) = 9.01 | < .05 | partial $\eta^2$ = .41 |
|  | guided- and free-movement behavior combined | main effect of test condition | F(1, 10) = 29.11 | < .001 |  |
|  |  | guided vs. free for 'audio-only' | F(1, 10) = 29.88 | < .001 | r = .87 |
|  |  | guided vs. free for video | F(1, 10) = 10.50 | < .01 | r = .72 |

### 3.2. Task 2: Speech intelligibility and localization

The movement behavior of the subjects during this task was analyzed according to the method described in section 2.7. Head-H and eye-movement trajectories were averaged over trials and subjects and grouped by visual condition and whether the location, color and number were answered correctly. The average movement trajectories are shown in Figure 6. Furthermore, the average head direction during the presentation of the name keyword was calculated for each visual condition, grouped by target speaker position, and compared to the optimal head direction for SNR in the better ear (Table 3). Finally, the average number of gaze jumps during each trial was calculated, grouped by visual condition (Fig. 7B).

**Movement trajectories.** The average gaze and head direction trajectories in azimuth (Fig. 6) show that most movement was of the eyes, not the head. In the animation conditions, subjects looked at what they thought was the target speaker. A clear difference between visual conditions can be seen: subjects moved directly towards the target speaker if the animated characters gazed at the target speaker, because before the name keyword was said they could already know who the target speaker was. In the other conditions, they moved towards the target speaker only after the name keyword was voiced. In the 'audio-only' condition, subjects on average did not move as closely to the target speaker location as in the animation conditions. In the 'audio-only' condition, most subjects did not move at all and looked and pointed their head at the center. However, some subjects did move if one of the edge speakers was the target speaker, as can be seen by the larger head-direction angle for the edge speakers in Table 3. Therefore the average gaze trajectory moved towards the target speaker, but stopped further away from it than in the other conditions.

**Initial difference in correctly and incorrectly answered trials.** At the beginning of each trial (0 s), there was a clear difference in average head- and eye direction relative to the target speaker between the correctly and incorrectly answered trials. This occurred because there was a difference in performance between trials where the middle or edge speakers were target speaker. Performance when the edge speakers were target speaker was on average 26.12% better than when the middle speakers were target speaker. Therefore, the average movement trace for the correctly answered trials contained more trials where the edge speaker was target speaker. This resulted in a larger angle relative to the target speaker at the beginning of the trial, considering that at the beginning of each trial, the subjects were mostly looking and pointing their head at the center. Furthermore, the head rotation when the subjects were looking at the edge speakers compared to the middle speakers was relatively smaller; because more of the movement was done with the eyes (see Fig. 2). Therefore, the head direction relative to the target speaker was bigger for the edge speakers. Combined with the better performance for the edge speakers, this explains why the average head direction trajectory for the correctly answered trials is further away from the target speaker location than the average head direction trajectory for the incorrectly answered trials.

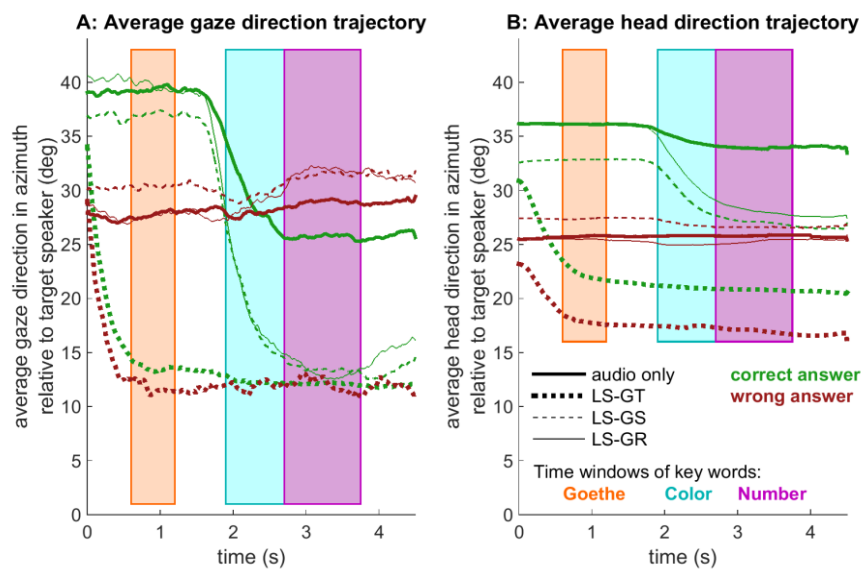

**Figure 6** Average movement trajectories during the CRM task. Shown are the gaze (A) and head (B) direction in azimuth over time relative to the target speaker location and averaged over all trials and all subjects (N=14). Trials where the location, color, and number were answered correctly (green lines) are shown separately from trials where one or more answers were incorrect (red lines). The visual conditions are plotted using different line styles: audio-only (N = 184 correct answer, N = 264 wrong answer), LS-GT (lip-sync, gaze to target, N = 384 correct answer, N = 64 wrong answer), LS-GS (lip-sync, gaze to subject, N = 216 correct answer, N = 232 wrong answer), LS-GR (lip-sync, gaze random, N = 198 correct answer, N = 250 wrong answer). The time windows for the key words are indicated using colored panels.

**Head direction relative to center.** The average head direction in azimuth relative to the center during the number key word (Table 3) shows that subjects generally pointed their heads towards the center. For the edge target speakers, subjects moved their head more towards the target speaker location in the animation conditions, especially in the 'lip-sync, gaze to target' condition. The head direction angles for the edge target speakers in the animation conditions are close to the optimal SNR head direction, but the angles for the middle target speakers are not.

**Table 3** The average head direction in azimuth relative to the center in the CRM task during the number key word compared to the optimal head direction for SNR in the better ear. Trials where one of the middle speakers was target speaker are separated from those where one of the edge speakers was target speaker and only correctly answered trials are included. For target speakers at negative angles, the sign of the head orientation was flipped. Optimal SNR head direction follows from model predictions for the SNR in the better ear.

| Target speaker position | Optimal SNR head direction relative to center | Head direction relative to center during number keyword | | | |
|---|---|---|---|---|---|
| | | Audio-only | Lip-sync, gaze to subject | Lip-sync, gaze to target | Lip-sync, gaze random |
| +15° | -30° | 3.72±6.20° | 1.75±2.96° | 4.38±3.55° | 1.49±3.20° |
| +45° | +15° | 2.29±3.33° | 8.64±5.36° | 16.09±7.69° | 10.22±6.57° |

**Effect of visual condition on task performance and gaze jumps.** As a measure of task performance, the percentage of trials in which the location, color and number were correctly answered was calculated, grouped by visual condition. The average of correct trials for the 'lip-sync, gaze to target' was 85.7% (std = 3.2%), whereas it was 41.1% (std = 4.5%), 48.2% (std = 3.3%) and 44.2% (std = 3.0%) for the conditions 'audio-only', 'lip-sync, gaze to subject' and 'lip-sync, gaze random', respectively (Fig. 7A). A repeated-measures ANOVA (values in Table 4) of the task performance (percentage of trials where the location, color, and number were correct) and the average number of gaze jumps during each trial was carried out. It shows that the visual condition has a significant effect on the task performance and on the average number of gaze jumps. Post-hoc tests with Bonferroni correction revealed that the task performance for the 'lip-sync, gaze to target' condition was significantly better than that for the other visual conditions (Fig. 7A). It also revealed that subjects made significantly fewer gaze jumps in the 'audio-only' condition (Fig. 7B), and significantly more gaze jumps in the 'lip-sync, gaze random' condition (Fig. 7B). There were no trials in which the color and number, but not the location, were correctly identified. A comparison of the task performance for the different target speakers shows that subjects had 59.4% and 49.1% of trials correct for the original male and female voices, respectively, and 46.9% and 63.8% of trials correct for the artificial male and female voices, respectively. Thus subjects had a similar performance with the artificial voices as for the original voices. Differences in performance between target speakers were found, but they did not influence the outcome of the task, as each target speaker was presented equally often.

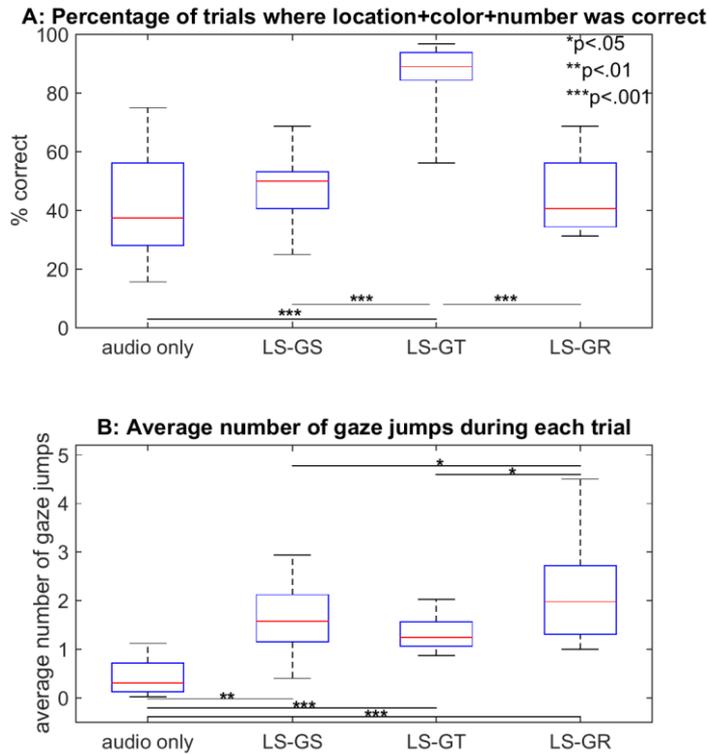

**Figure 7** Boxplots (range, interquartile and median, outliers are plotted as red crosses) for the CRM task of A, Percentage of trials where the location, color and number were identified correctly, and B, The average number of gaze jumps during each trial. The visual conditions are indicated in the figure as: audio-only, LS-GS (lip-sync, gaze to subject), LG-GT (lip-sync, gaze to target), LS-GR (lip-sync, gaze random). Boxplots show the averages for 14 subjects over 32 trials.

**Table 4 Statistics results Task 2.**

| Measure | Comparison | F | p | Effect size |
|---|---|---|---|---|
| task performance | main effect of visual condition | F(3, 39) = 72.14 | < .001 | partial $\eta^2$ = .85 |
| | 'lip-sync, gaze to target' vs. 'audio-only' | | < .001 | |
| | 'lip-sync, gaze to target' vs. 'lip-sync, gaze to subject' | | < .001 | |
| | 'lip-sync, gaze to target' vs. 'lip-sync, gaze random' | | < .001 | |
| average number of gaze jumps | main effect of visual condition | F(1.72, 22.30) = 21.66* | < .001 | partial $\eta^2$ = .63 |
| | *Mauchly's test was significant: $\chi^2$ = 13.22, p < .05, correction with Greenhouse-Geisser estimate of sphericity $\varepsilon$ = .57* | | | |
| | 'audio-only' vs. 'lip-sync, gaze to subject' | | < .01 | |
| | 'audio-only' vs. 'lip-sync, gaze random' | | < .001 | |
| | 'audio-only' vs. 'lip-sync, gaze to target' | | < .001 | |

| | 'lip-sync, gaze to target' vs. 'lip-sync, gaze to subject' | | < .05 | |
| | 'lip-sync, gaze to target' vs. 'lip-sync, gaze random' | | < .05 | |

## 3.3. Task 3: Subjective rating of the visual conditions

The subjects' responses to the questionnaire are plotted in Figure 8. A repeated-measures ANOVA (values in Table 5) of the ratings shows that there is a significant effect of visual condition on the ratings of general impression, perceived speech intelligibility, and perceived listening effort. Contrasts between the video condition and the other conditions revealed that subjects rated the video condition with a general impression significantly better than all the other conditions (Fig. 7 Q1). The 'lip-sync, gaze to target' condition was rated the best of all animation conditions, with a score of "normal – very good" on average. Moreover, the 'lip-sync, gaze to target' condition did not significantly differ from the video condition in the ratings for perceived speech intelligibility (Fig. 7 Q2) and perceived listening effort (Fig. 7 Q3). No effect of visual conditions on the perceived audio quality was found.

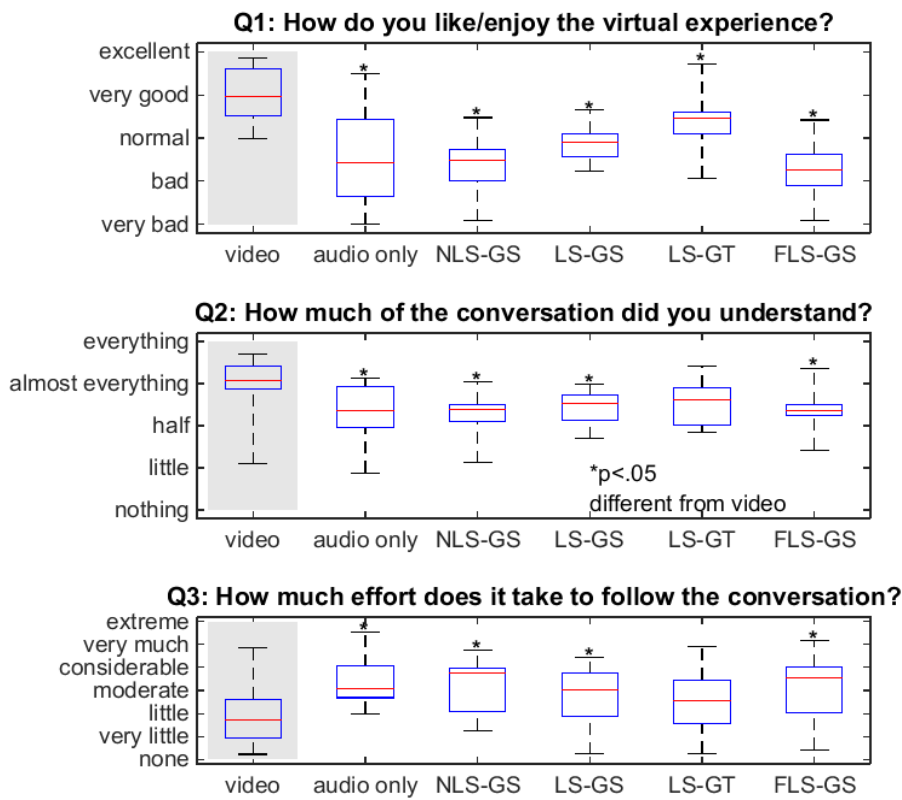

**Figure 8 Boxplots** (range, interquartile and median) of the responses (N=14) for the rating task on the questions about the general impression (Q1), perceived speech intelligibility (Q2) and perceived listening effort (Q3). The visual conditions are indicated in the figure as: video, audio-only, NLS-GS (no lip-sync, gaze to subject), LS-GS (lip-sync, gaze to subject), LG-GT (lip-sync, gaze to target), FLS-GS (fish mouth, gaze to subject). Visual conditions that differ significantly from the video condition are marked.

Table 5 Statistics results Task 3.

| Measure | Comparison | F | p | Effect size |
|---|---|---|---|---|
| general impression | main effect of visual condition | F(2.41, 31.26) = 17.48* | < .001 | |
| | *Mauchly's test was significant: $\chi^2(14)$ = 31.93, p < .01, correction with Greenhouse-Geisser estimate of sphericity ε = .48* | | | |
| | video vs. 'audio-only' | F(1,13) = 15.37 | < .01 | partial $\eta^2$ = .54 |
| | video vs. 'no lip-sync, gaze to subject' | F(1,13) = 37.07 | < .001 | partial $\eta^2$ = .74 |
| | video vs. 'lip-sync, gaze to subject' | F(1,13) = 23.35 | < .001 | partial $\eta^2$ = .64 |
| | video vs. 'lip-sync, gaze to target' | F(1,13) = 5.62 | < .05 | partial $\eta^2$ = .34 |
| | video vs. 'fish mouth, gaze to subject' | F(1,13) = 36.68 | < .001 | partial $\eta^2$ = .74 |
| perceived speech intelligibility | main effect of visual condition | F(5, 65) = 3.97 | < .01 | |
| | video vs. 'audio-only' | F(1,13) = 9.72 | < .01 | partial $\eta^2$ = .43 |
| | video vs. 'no lip-sync, gaze to subject' | F(1,13) = 22.45 | < .001 | partial $\eta^2$ = .63 |
| | video vs. 'lip-sync, gaze to subject' | F(1,13) = 8.45 | < .05 | partial $\eta^2$ = .39 |
| | video vs. 'lip-sync, gaze to target' | F(1,13) = 4.63 | .051 | partial $\eta^2$ = .26 |
| | video vs. 'fish mouth, gaze to subject' | F(1,13) = 6.77 | < .05 | partial $\eta^2$ = .34 |
| perceived listening effort | main effect of visual condition | F(2.75, 35.70) = 6.32* | < .01 | |
| | *Mauchly's test was significant: $\chi^2(14)$ = 25.63, p < .05, correction with Greenhouse-Geisser estimate of sphericity ε = .55* | | | |
| | video vs. 'audio-only' | F(1,13) = 11.23 | < .01 | partial $\eta^2$ = .46 |
| | video vs. 'no lip-sync, gaze to subject' | F(1,13) = 17.72 | < .01 | partial $\eta^2$ = .58 |
| | video vs. 'lip-sync, gaze to subject' | F(1,13) = 5.14 | < .05 | partial $\eta^2$ = .28 |
| | video vs. 'lip-sync, gaze to target' | F(1,13) = 4.03 | .066 | partial $\eta^2$ = .24 |
| | video vs. 'fish mouth, gaze to subject' | F(1,13) = 17.97 | < .001 | partial $\eta^2$ = .58 |

## 4. Discussion

### 4.1. Effect of visual cues on movement behavior

Analysis of the head and eye movement behavior in the listening to conversations and CRM tasks shows that the movement behavior was influenced by the visual condition. In the 'audio-only' and 'no lip-sync, gaze to subject' conditions, the movement behavior was significantly different from the movement behavior in the video condition (Fig. 4). These findings are in correspondence with our hypotheses that speech-driven lip-syncing was necessary to obtain similar movement behavior as in the video condition, except that the 'fish mouth, gaze to subject' condition was also expected to be significantly different. In the following, the movement strategies used by the subjects are discussed for each group of conditions that were hypothesized to have the same movement strategy.

**Audio-only.** It was hypothesized that in the audio-only condition, subjects would follow an SNR-optimization or a localization-based movement strategy. Because most subjects did not move at all in the 'audio-only' condition, there is no evidence for a localization-based movement strategy in this condition. The SNR for the individual head orientations was higher in the 'audio-only' condition, but with a very large variance (Fig. 4D). Always looking at the center would also result in a higher SNR, because one of the ears is pointed more towards the target source than it would be when looking at the target. There is thus no evidence that young, normal-hearing subjects actively move their head to optimize SNR. This is in accordance with the findings of Grange & Culling (2016).

**'fish mouth' and 'no lip-syncing'.** In the 'fish mouth, gaze to subject' and 'no lip-sync, gaze to subject' conditions, looking at the active speaker provides no speech-intelligibility benefit. It was therefore expected that subjects would have an SNR-optimization based strategy. However, in the 'fish mouth, gaze to subject' condition, subjects had the same movement behavior of looking at the active speaker as in the video condition. An explanation could be that subjects were expecting useful information to be present in the lip-syncing and therefore kept looking at it. In the 'no lip-sync, gaze to subject' condition, the movement behavior was significantly different from that in the video condition. Subjects were able to identify the new speaker after a speaker change in the 'no lip-sync, gaze to subject' condition as fast as in the video condition (Fig. 4C) and were also looking at the active speaker, but for a smaller percentage of the time. However, they were slower to identify the new speaker after a speaker change in the 'audio-only' condition for the guided movement behavior (Fig. 5B). It is surprising that there is a difference between those two conditions, because in both conditions there was no lip-syncing that could help identify who was talking. A possible explanation could be that subjects can predict from the conversation content who is going to talk next and that they can more quickly shift their gaze towards this person because they see where he/she is located, which they cannot do in the 'audio-only' condition. The purpose of looking at the active speaker for a smaller percentage of time could not have been to optimize the SNR, because the estimated SNR for the individual head orientations was not better than in the video or other animation conditions (Fig. 4D).

**Video and animation conditions with speech-driven lip-syncing.** It was hypothesized that in the video condition and animation conditions with speech-driven lip-syncing, subjects would follow the movement strategy of looking at the active speaker. The results show that subjects looked at the active speaker in these conditions, although they did not do this as accurately as when specifically instructed to do so (guided movement behavior) (Fig. 5). Vertegaal et al. (2001) showed that during natural conversations, people look at the individual they listen to only 62% of the time, which could be due to social interaction habits, i.e., how people in a conversation take turns and look at each other. In our study, the subjects did not socially interact, because no real persons were involved in the reproduction of the conversation, and the subjects were not participating in it themselves. However, subjects may still have sought non-verbal information on other participants of the conversation, showing, e.g., a facial expression in reaction of a comment or giving head nods to show understanding, and thus occasionally turning away from the active speaker. Further studies comparing real-life and audiovisual virtual environments are required to clarify this issue. Occasionally looking away from the active speaker, and thereby pointing the head in a different direction, increases the SNR, as is shown by the comparison with the guided-movement behavior (Fig. 5). Therefore, looking at the active speaker for a smaller percentage of the time, and thereby pointing the head to the active speaker for a smaller percentage of time, could also have been part of an SNR optimization strategy. However, that this is unlikely becomes clear when looking at the CRM task. The average head direction angles relative to the center during the number keyword for the edge target speakers in the animation conditions are close to the optimal SNR head direction, but the angles for the middle target speakers are not. Therefore, the question arises whether the head direction angles are rather a consequence of looking at the target speaker than a consequence of SNR optimization. To answer this, we looked at values for head-eye angle relationships found in literature. If the gaze direction is towards the target speaker, the head direction will only be part of this gaze angle (Fig. 2). The head direction angles resulting from looking at the target speaker (13.3° and 29.6° for gaze directions of 15° and 45°, respectively (Uemura et al. 1980)) fit the measured head direction angles shown in Table 3 better than the angles for the optimal SNR head direction. Furthermore, in Figure 6A we can see that there was actually a shift of gaze direction towards the target speaker during the CRM task. Thus it is more likely that the subjects were looking at the target speaker than that they were actively moving their head to optimize SNR.

### 4.2. Effect of visual cues on speech intelligibility and localization task performance

Analysis of the CRM data shows that task performance was influenced by the visual condition. In the 'lip-sync, gaze to target' condition, the SNR was the same as in the other conditions, but the percentage of trials where the location, color and number were answered correctly was higher (Fig. 7A).. A possible reason for this increased task performance could be that the gaze direction helps identify which speaker to attend to. As previously shown by Kidd et al. (2005), knowing where to attend to results in a better task performance. Another study by Teraoka et al. (2017) analyzed the effect of spatial attention on speech intelligibility by comparing word intelligibility for target speech presented at the expected or at an unexpected location and showed that the effects of auditory spatial attention might affect word intelligibility. The results of the current study and of the studies by Kidd et al. and Teraoka et al. show that attention is a very important factor for speech intelligibility, and might even be more important than the small SNR improvement that can be achieved by head orientation. The results of the current study also indicate that attention can be guided by visual cues, which argues for the inclusion of visual cues in future experiments. Especially in listening tasks that involve multiple sources, inclusion of visual cues is important, because the visual cues can guide attention to the correct source as shown in this study. There were no trials where the color and number, but not the location, were identified correctly. This shows that localization of the target is important for speech comprehension, as shown by Yost et al. (1996). It is likely that subjects do not know where to steer their auditory spatial attention if they cannot localize the target. The significantly higher average number of gaze jumps during each trial (Fig. 7B) is an indication that the subjects were distracted by the random gaze direction of the animated characters.

### 4.3. Effect of visual cues on subjective rating

It is not surprising that the video condition was given the best general impression rating (Fig. 8), because there are behavioral features in the video that are important for speech comprehension but were not implemented for the animated characters, such as facial expressions and gestures. In addition, the gaze patterns could be made more natural and the lip-syncing could be improved. However, the rating for the 'lip-sync, gaze to target' condition was also good and no significant differences in perceived speech intelligibility and perceived listening effort were reported compared to the video condition , although these differences wereonly borderline non-significant, and a larger sample size might result in significance. Furthermore, the contribution of the speech-driven lip-syncing to the speech intelligibility task performance was not analyzed in this study. Therefore, further research with more elaborate speech intelligibility tests is required to validate the animated characters in terms of speech intelligibility. However, as expected, the results do indicate that the subjects felt comfortable with the animated characters in the 'lip-sync, gaze to target' condition. Moreover, adding lip-syncing and gaze direction towards the active speaker was preferred over virtual characters without lip-syncing and with static gaze direction towards the subject. This confirms the hypothesis that the 'lip-sync, gaze to target' condition would receive the best subjective ratings of the animation conditions.

### 4.4. Building blocks for audiovisual environments

So far, we determined visual factors that influenced the head-and eye-movement behavior, and combined localization and speech intelligibility task performance and perceived speech intelligibility, perceived listening effort and the general impression of the audiovisual environments. However, our long-term goal is to provide building blocks for audiovisual environments that foster natural movement behavior, where ecologically valid task performance can be measured and with which normal hearing subjects and hearing aid users are comfortable. Here future steps are discussed that need to be taken in order to reach this long-term goal.

**Towards ecological validity of the movement behavior.** Results show that animated characters with lip-syncing (speech-driven or 'fish mouth') are necessary for subjects to express the same movement behavior as in the video condition. Video recordings of real persons projected onto a cylindrical screen are the most ecologically valid condition that can be achieved in the laboratory used in this experiment. Advanced technologies such as 3D recordings with light-field cameras generate a higher level of immersion when displaying the 3D recordings via head-mounted displays or stereo displays. However, these technologies are expensive and complex and have not been validated in terms of head movement. We are aware that movement behavior under the video condition might not be fully ecologically valid, and a comparison of behavior in two audiovisual environments to behavior in real life is currently being made in a separate study (Paluch et al. 2018). In addition, a study comparing behavior when visual stimuli are displayed on a head-mounted display and projected onto a cylindrical screen is in progress. Because our reference condition (video) might not be fully ecologically valid, we cannot claim ecological validity of the animation conditions that induce similar movement behavior. However, showing that the movement behavior under the animation conditions with lip-syncing is similar to that in the most realistic condition that can be reproducibly achieved in the lab is an important step towards ecological validity.

**Towards ecological validity of the task performance.** The combined localization and speech intelligibility task shows that visual cues affect performance in specific tasks. To assess the ecological validity of the animated characters regarding speech intelligibility performance, further research with more elaborate speech intelligibility tests and a comparison to video material of real persons is required. Most speech intelligibility tests (including the CRM task) are not ecologically valid themselves, containing unnatural speech and noise. Therefore it makes sense to use more realistic speech tests, such as speech comprehension tests (e.g., Best et al. 2016), to test ecological validity in future studies. If the speech intelligibility or speech comprehension benefit of the speech-driven lip-syncing differs too much from the benefit of the video material, improvement of the lip-syncing method would be necessary.

The subjective rating shows that the subjects are comfortable with the animated characters in the 'lip-sync, gaze to target' condition. Planned studies involve visual scenes with not only animated characters, but also animated background, such as a scene showing animated characters sitting in a cafeteria. In studies using such visual scenes, it is important how life-like these scenes appear to the subjects and whether they feel that they are immersed in the scene. Therefore, these planned studies will involve more elaborate questionnaires.

As a first step, only young, normal-hearing participants were included in the current study. Older people and hearing-aid users could exhibit different head and eye movement behavior and might react differently to the animated characters. Further research on the behavior of older people and hearing-aid users and their reaction to the animated characters is required.

## 5. Conclusion and Outlook

In this study, the influence of visual cues on head and eye movement, on combined speech intelligibility and localization task performance, as well as on subjective ratings of perceived speech intelligibility, perceived listening effort and the general impression of the audiovisual environments was investigated. Animated characters with different lip-syncing and gaze patterns were compared to an audio-only condition and to a video of real persons in an attempt to find a level of realism of the animated characters at which the movement behavior of the subjects was similar to that in the video condition.

**Visual cues affect head and eye movement.** If visual cues with lip-syncing (animations and video) are present, subjects look at the active speaker most of the time or, in the case of the combined localization and speech intelligibility task, look at what they think is the target speaker. The presence of (animated) speakers, even if they have no lip-syncing and a fixed gaze direction towards the subject, causes a faster gaze shift towards the new speaker after a speaker change. If no visual cues are present, most subjects do not move at all, but always look at the center. No evidence was found that subjects actively try to optimize signal-to-noise ratio using head movement, or that they have a localization-based head movement strategy. Even if there was no benefit in looking at the active speaker (animation condition without lip-syncing and gaze direction of the animated characters towards the subject), subjects still did this, but for a smaller percentage of time than in the video condition.

**Visual cues affect task performance.** Combined localization and speech intelligibility task performance increases if the animated characters gaze towards the target speaker, because subjects then know where to attend to.

**Visual cues affect subjective ratings.** Subjects reported less perceived listening effort and better perceived speech intelligibility for video or animated characters with speech-driven lip-syncing and gaze to the target speaker. Subjects reported the best general impression for the video condition, followed by the animation condition with speech-driven lip-syncing and gaze direction towards the target speaker.

**Towards ecological validity.** For measuring movement behavior similar to that in the video condition, animated characters with any type of lip-syncing are sufficient. For task performance, gaze direction is also important. Subjects felt comfortable with the most realistic animation condition, with speech-driven lip-syncing and gaze direction towards the target speaker, and preferred this condition over animated characters without lip-syncing and with static gaze direction towards the subject. We conclude that we can use this most realistic animation condition for measuring movement behavior in young, normal-hearing listeners instead of the video condition. This contributes to our long-term goal of finding a level of realism of animated characters at which users move in an ecologically valid way, at which their task performance is also ecologically valid and with which they are comfortable.

Based on the work presented here, further studies will investigate the effect of hearing-aid signal processing on motion behavior and task performance in elderly and hearing-impaired listeners, with the aim of identifying limitations of current technology in more ecologically valid environments. In parallel, the contribution to speech intelligibility of the animated lip-syncing method employed here will be evaluated in more detail in comparison to video recordings.

## 6. Acknowledgements

This study was funded by DFG research unit FOR1732 "Hearing Acoustics" and by European Union's Horizon 2020 research and innovation programme under the Marie Sklodowska-Curie grant agreement No 675324 (ENRICH). We thank J. Rieger and I. Schepers for providing us with the German version of the CRM corpus. We also thank K. Schwarte for her help with the measurements and J. Luberadzka, C. Hauth and T. Gerdes for their role as actors in the conversations. We also thank the three anonymous reviewers for their helpful comments and suggestions. Language services were provided by STELS-OL.de.